\documentclass[12pt,a4paper]{article}
\usepackage{a4wide,cite}

\newcommand{\be}{\begin{equation}}
\newcommand{\ee}{\end{equation}}
\newcommand{\bea}{\begin{eqnarray}}
\newcommand{\eea}{\end{eqnarray}}
\newcommand{\ep}{\varepsilon}

\newcommand{\nn}{\nonumber}

\catcode`\@=11

\begin{document}


 \thispagestyle{empty}
 \begin{flushright}
 {MZ-TH/99-30} \\[3mm]
 {hep-ph/9908032} \\[3mm]
 {August 1999}
 \end{flushright}
 \vspace*{2.0cm}
 \begin{center}
 {\Large \bf
 Geometrical approach to Feynman integrals \\[2mm]
 and the $\ep$-expansion\footnote{Contribution to the Proceedings of
International Workshop ``New Computing Techniques in Physics Research''
(AIHENP-99), Heraklion, Crete, Greece, 12--16 April 1999.}
}
 \end{center}
 \vspace{1cm}
 \begin{center}
 A.I.~Davydychev\footnote{ Alexander von Humboldt fellow. On leave from
                 Institute for Nuclear Physics, Moscow State University,
                 119899 Moscow, Russia. 
Email: davyd@thep.physik.uni-mainz.de}
\\
 \vspace{1cm}
{\em
 Department of Physics,
 University of Mainz, \\
 Staudingerweg 7,  
 D-55099 Mainz, Germany}
\end{center}
 \hspace{3in}
\begin{abstract}
Application of the geometrically-inspired representations
to the $\ep$-expansion of the two-point function with different
masses is considered.
Explicit result for an arbitrary term of the expansion
is obtained in terms of log-sine integrals.
Construction of the $\ep$-expansion in the three-point case
is also discussed. 
\end{abstract}


\newpage

{\bf 1.}
When studying Feynman diagrams, especially those with three or more 
external particles, one of the most important issues is 
how to choose appropriate variables.
In particular, these variables should be chosen in such a way that the
solutions of Landau equations \cite{Landau}, i.e.\ the positions of
possible singularities, could be easily identified and understood.
Studying the singularities of diagrams was the first reason of
introducing geometrical ideas for the three-point \cite{KW}
and four-point \cite{Wu} functions at one loop.
With a one-loop $N$-point diagram one can associate an $N$-dimensional
Euclidean basic simplex (for details, see e.g.\ in \cite{DD})
whose sides are directly related to the masses and the
external momenta invariants. Then, the situations when the volume 
(content) of this simplex vanishes correspond to the positions 
of possible singularities.
For instance, for a two-point function with an external momentum
$k_{12}$ this basic simplex is just a triangle whose two sides
are equal to $m_1$ and $m_2$ (the masses of the internal particles)
and the third side is $K_{12}\equiv\sqrt{k_{12}^2}$.
Its area vanishes when $k_{12}^2=(m_1\pm m_2)^2$, i.e.\ at the
threshold and pseudothreshold. 

Another important step was to understand that the evaluation of 
Feynman integrals themselves, not only looking for the positions 
of the singularities, could be also reduced to a purely geometrical 
problem. In particular, as it is shown in \cite{DD} 
(see also in \cite{DD-APP} for a brief review),
the Feynman parametric representation of a one-loop $N$-point integral 
can be directly transformed into an integral over one of the 
$N$-dimensional solid angles of this basic simplex. The latter can 
be understood as the integral over an $(N-1)$-dimensional 
simplex in non-Euclidean space of constant curvature. In particular, 
the four-point
function in four dimensions is proportional to the volume of 
a non-Euclidean tetrahedron in spherical or hyperbolic space. 
The three-point function in three dimensions is just the area of
a spherical (or hyperbolic) triangle; in this way, a very nice result
of \cite{Nickel} was reproduced in a purely geometrical way.
The three-point function in four dimensions is also an integral over
the same triangle, but with an extra weight factor $1/\cos\theta$,
where $\theta$ is the angle between the running vector of integration
and the direction of the height of the basic simplex (which is 
a tetrahedron in this case). One can even get rid of this weight factor,
reducing the three-point function to a special (asymptotic) case
of the four-point function \cite{Wagner}. 
For some other links 
between Feynman diagrams and non-Euclidean geometry, see in \cite{BB}.

Instead of repeating all conclusions given in \cite{DD}, we
have decided to present here an extended discussion of using the 
geometrically-inspired representations for the evaluation of integrals
in the framework of dimensional regularization \cite{dimreg},
when the space-time dimension is $n=4-2\ep$.
Sometimes it is possible to present results valid for an
arbitrary $\ep$, usually in terms of hypergeometric functions.
However, for practical purposes the coefficients
of the expansion in $\ep$ are important. 
In particular, in multiloop calculations higher terms
of the $\ep$-expansion of the one-loop functions are needed, since 
one can get contributions where these functions are
multiplied by singular factors containing poles in $\ep$.
In this paper, we derive an explicit result for an arbitrary
term of the $\ep$-expansion of two-point function with arbitrary
masses. We also discuss the construction of the $\ep$-expansion 
of the three-point function.

\vspace{3mm}

{\bf 2.}
According to section~IV of \cite{DD}, 
for the two-point function with an external
momentum $k_{12}$ and internal masses $m_1$ and $m_2$,
with unit powers of propagators ($\nu_1=\nu_2=1$), 
in $n=4-2\ep$ dimensions, we get
\begin{equation}
\label{J^2(4,1,1)}
J^{(2)}(4-2\varepsilon;1,1)
=\mbox{i}\;\pi^{2-\varepsilon}\;\Gamma(\varepsilon)\;
\frac{m_0^{1-2\varepsilon}}{\sqrt{k_{12}^2}} \;
\left\{ \Omega_1^{(2; 4-2\varepsilon)}
+ \Omega_2^{(2; 4-2\varepsilon)} \right\} ,
\end{equation}
where
\begin{equation}
\label{Omega_i^(2;4-2ep)}
\Omega_i^{(2; 4-2\varepsilon)} =
\int_0^{\tau_{0i}}
\frac{\mbox{d}\theta}{\cos^{2-2\varepsilon}\theta} .
\end{equation}
Here it is assumed that $(m_1-m_2)^2\leq k_{12}^2 \leq (m_1+m_2)^2$.
In other regions one needs to use analytic continuation. 
The following notation \cite{DD} is used:
\be  
\label{two-point}
\cos\tau_{12} = \frac{m_1^2+m_2^2-k_{12}^2}{2m_1 m_2}, \hspace{5mm}
m_0 = \frac{m_1 m_2 \sin\tau_{12}}{\sqrt{k_{12}^2}}, 
\hspace{5mm}
\cos\tau_{0i} = \frac{m_0}{m_i} \; .
\ee
Note that
$\tau_{01}+\tau_{02} = \tau_{12}$ and
\be
\label{tan(tau0i)}
\tan\tau_{01}=\frac{m_1^2-m_2^2+k_{12}^2}{2m_0\sqrt{k_{12}^2}}, 
\hspace{5mm}
\tan\tau_{02}=\frac{m_2^2-m_1^2+k_{12}^2}{2m_0\sqrt{k_{12}^2}}.
\ee
The integral (\ref{Omega_i^(2;4-2ep)}) can be presented
in terms of the hypergeometric function 
$_2F_1$ given in eq.~(4.12) of \cite{DD} (see also in \cite{BDS}).
Expanding (\ref{Omega_i^(2;4-2ep)}) in $\ep$, we get
\be
\label{ep-exp}
\int_0^{\tau}
\frac{\mbox{d}\theta}{\cos^{2-2\ep}\theta}
= \sum\limits_{j=0}^{\infty} \frac{(2\ep)^j}{j!}
\int_0^{\tau}
\frac{\mbox{d}\theta}{\cos^2\theta}
\ln^j(\cos\theta) 
= \sum\limits_{j=0}^{\infty} \frac{(-2\ep)^j}{j!}
f_j(\tau),
\ee
\be
\label{f_j}
f_j(\tau) \equiv (-1)^j
\int_0^{\tau}
\frac{\mbox{d}\theta}{\cos^2\theta}
\ln^j(\cos\theta) .
\ee
The lowest terms of the expansion are 
(cf.\ eqs.~(4.9) and (4.11) of \cite{DD})
\bea
f_0(\tau) &=& \tan\tau, \\
\label{f_1}
f_1(\tau) &=& -\tan\tau \ln(\cos\tau)-\tan\tau +\tau, \\
\label{f_2}
f_2(\tau) &=& 
\tan\tau \left[\ln^2(\cos\tau)\!+\!2\ln(\cos\tau)\!+\!2 \right]
-2\tau (1\!-\!\ln 2) - \mbox{Cl}_2\left(\pi\!-\!2\tau\right) .
\hspace{4mm}
\eea 

Integrating by parts, we see that the $f_j$ function (\ref{f_j})
can be reduced to the generalized Lobachevsky function  
\be
\label{L_j}
L_j(\tau)\equiv (-1)^{j-1} \int_0^{\tau}
\mbox{d}\theta \; \ln^{j-1}(\cos\theta)
\ee
via
\be
f_j(\tau)= (-1)^j \tan\tau \ln^j(\cos\tau)
+ j L_j(\tau)- j f_{j-1}(\tau) .
\ee
Repeating this procedure, we arrive at 
\be
\label{f_j-expl}
f_j(\tau)=
j! \; \sum\limits_{l=1}^j \frac{(-1)^{j-l}}{(l-1)!}L_l(\tau) 
+(-1)^j \; j! \; \tan\tau \; \sum\limits_{l=0}^j
\frac{\ln^l(\cos\tau)}{l!} .
\ee

When we consider the infinite sum over $j$, eq.~(\ref{ep-exp}),
the logarithmic tail in eq.~(\ref{f_j-expl}) can be summed up
into a closed form, since
\be
\sum\limits_{j=0}^{\infty} (2\ep)^j 
\sum\limits_{l=0}^j 
\frac{\ln^l(\cos\tau)}{l!}
= \frac{\cos^{2\ep}\tau}{1-2\ep} .
\ee
Therefore,
\be
\label{ep-exp2}
\int_0^{\tau}
\frac{\mbox{d}\theta}{\cos^{2-2\ep}\theta}
=\frac{\cos^{2\ep}\tau}{1-2\ep} \; \tan\tau
+ \sum\limits_{j=1}^{\infty} (2\ep)^j 
\sum\limits_{l=1}^j \frac{(-1)^l}{(l-1)!}\; L_l(\tau) .
\ee

It is obvious that $L_1(\tau)=\tau$. Then,
note that $L_2(\tau)\equiv L(\tau)$ is what is usually
called the Lobachevsky function (see e.g.\ in Appendix~A 
of \cite{DD}). It is related to the Clausen function 
$\mbox{Cl}_2$ via
$L_2(\tau)= -\textstyle{1\over2}\;
\mbox{Cl}_2\left(\pi-2\tau\right) + \tau \ln 2$.

By a simple transformation of the integration variable,
the $L_j$ function (\ref{L_j})
can be presented as a finite sum of log-sine integrals
(see in \cite{Lewin}, chapter~7.9)
\be
\label{Ls_j}
\mbox{Ls}_j(\theta)= - \int_0^{\theta}
\mbox{d}\theta' \ln^{j-1}\left|2\sin\frac{\theta'}{2}\right|, 
\ee
namely:
\be
\label{L-Ls}
L_j(\tau)=\frac{(j-1)!}{2} \sum\limits_{i=1}^j
\frac{(-1)^i \; \ln^{j-i}2}{(j-i)! \; (i-1)!}
\left[ \mbox{Ls}_i(\pi) - \mbox{Ls}_i(\pi-2\tau) \right] .
\ee
Note that $\mbox{Ls}_1(\theta)=-\theta$,  
$\mbox{Ls}_2(\theta)=\mbox{Cl}_2(\theta)$,
$\mbox{Ls}_2(\pi)=0$. Some other values of $\mbox{Ls}_i(\pi)$
are given in eq.~(7.113) of \cite{Lewin} in terms of $\zeta(j)$.
The result for $\mbox{Ls}_i(\pi)$ with 
an arbitrary $i$ is given in eq.~(45) of Appendix~A.2.7
of \cite{Lewin} in terms of a differential operator, which can 
be represented in a more compact form as
\be
\label{diff}
\mbox{Ls}_{j+1}(\pi)= -\frac{\pi}{2^j} 
\left. \left(\frac{\mbox{d}}{\mbox{d}z}\right)^j
\frac{\Gamma(1+2z)}{\Gamma^2(1+z)}\right|_{z=0} .
\ee

For $j=3$, eq.~(\ref{L-Ls}) yields
\bea
L_3(\tau)&=&\frac{1}{2} \; \mbox{Ls}_3(\pi-2\tau)
-\ln 2 \; \mbox{Cl}_2(\pi-2\tau)
+\tau \ln^2 2 + \frac{\pi^3}{24} ,
\nn \\
f_3(\tau) &=& -\tan\tau \left( \ln^3(\cos\tau) 
+ 3\ln^2(\cos\tau) + 6\ln(\cos\tau) + 6 \right)
\nn \\
&& + 3\tau \left( 2 - 2\ln 2 + \ln^2 2 \right)
\nn \\
&& + \frac{\pi^3}{8}
+ 3 ( 1-\ln 2 ) \; \mbox{Cl}_2(\pi-2\tau)
+\frac{3}{2}\; \mbox{Ls}_3(\pi-2\tau) .
\hspace{6mm}
\eea
Note that the function $\mbox{Ls}_3$ has also occurred in the 
$\ep$-part of a triangle diagram with massless internal
particles \cite{DT2}.

Substituting (\ref{L-Ls}) into eq.~(\ref{ep-exp2}), 
the resulting three-fold sum can be simplified,
\bea
\frac{1}{2} \sum\limits_{j=1}^{\infty} (2\ep)^j 
\sum\limits_{l=1}^j \sum\limits_{i=1}^l 
\frac{(-1)^{l-i} \ln^{l-i}2}{(l-i)! (i-1)!}
\left[ \mbox{Ls}_i(\pi) - \mbox{Ls}_i(\pi-2\tau) \right]
\hspace{20mm}
\nn \\
= \frac{2^{-2\ep}\; \ep}{1-2\ep}
\sum\limits_{j=0}^{\infty} \frac{(2\ep)^j}{j!}
\left[ \mbox{Ls}_{j+1}(\pi) - \mbox{Ls}_{j+1}(\pi-2\tau) \right] .
\eea
Therefore,
\bea
\label{ep-exp3}
\int_0^{\tau}
\frac{\mbox{d}\theta}{\cos^{2-2\ep}\theta}
= \frac{\cos^{2\ep}\tau}{1-2\ep} \tan\tau
+\frac{2^{-2\ep} \ep}{1\!-\!2\ep}
\sum\limits_{j=0}^{\infty} \frac{(2\ep)^j}{j!}
\left[ \mbox{Ls}_{j+1}(\pi) 
- \mbox{Ls}_{j+1}(\pi\!-\!2\tau) \right] .
\nn \\
{}
\eea
Finally, 
taking into account eqs.~(\ref{two-point})--(\ref{tan(tau0i)}),
we obtain for the integral (\ref{J^2(4,1,1)})
\bea
\label{2pt_res2}
J^{(2)}(4-2\ep;1,1) &=& \mbox{i}\pi^{2-\ep} 
\frac{\Gamma(1\!+\!\ep)}{1-2\ep}
\left\{ \frac{m_1^{-2\ep} \!+\! m_2^{-2\ep}}{2\ep}
+ \frac{m_1^2\!-\!m_2^2}{2\ep \; k_{12}^2} 
\left( m_1^{-2\ep} \!-\! m_2^{-2\ep} \right)
\right.
\nn \\
&& 
+ \frac{2^{-2\ep}(m_1 m_2 \sin\tau_{12})^{1-2\ep}}{(k_{12}^2)^{1-\ep}}
\sum\limits_{j=0}^{\infty} \frac{(2\ep)^j}{j!}
\nn \\
&& \left. \times
\left[ 2\mbox{Ls}_{j+1}(\pi)\!
-\! \mbox{Ls}_{j+1}(\pi\!-\!2\tau_{01})\!
-\! \mbox{Ls}_{j+1}(\pi\!-\!2\tau_{02})
\right] \frac{}{} \right\} .
\hspace{3mm}
\eea
The coefficient at $\ep^j$ has a closed form in terms
of the $\mbox{Ls}_{j+1}$ functions, whose arguments have a simple
geometrical interpretation.
Note that, according to eq.~(\ref{diff}), the infinite sum with 
$\mbox{Ls}_{j+1}(\pi)$ in eqs.~(\ref{ep-exp3})--(\ref{2pt_res2})
can be converted into 
\be
\sum\limits_{j=0}^{\infty} \frac{(2\ep)^j}{j!}\; 
\mbox{Ls}_{j+1}(\pi) = -\pi\; \frac{\Gamma(1+2\ep)}{\Gamma^2(1+\ep)} \; .
\ee

As it was mentioned before, the obtained result (\ref{2pt_res2})
can be directly applied in the region 
$(m_1-m_2)^2\leq k_{12}^2 \leq (m_1+m_2)^2$.
In other regions, the analytic continuation of (\ref{2pt_res2})
gives expressions in terms of (generalized)
polylogarithms. The result for the $\ep$-term
was obtained in \cite{NMB}.  
For the case $m_1=0$ ($m_2\equiv m$), the
first terms of the expansion (up to $\ep^3$) are presented
in eq.~(A.3) of \cite{FJTV}.

An important special case of eq.~(\ref{2pt_res2}) is
$m_1=m_2\equiv m$, $k_{12}^2=m^2$. In this case,
$\tau_{12}=\textstyle{{1\over3}}\pi$ and 
$\tau_{01}=\tau_{02}={\textstyle{1\over6}}\pi$.
Therefore, the $j$-th term of the expansion contains
$\mbox{Ls}_{j+1}(2\pi/3)$. In particular, this is 
the reason why $\mbox{Ls}_3(2\pi/3)$
appears in the results for some two-loop 
diagrams considered in \cite{FKK}.

\vspace{3mm}

{\bf 3.} 
The geometrical approach to the
three-point function is discussed in section~V of \cite{DD}.
The integration extends over a spherical (or hyperbolic)
triangle, as shown in Fig.~6 of \cite{DD},
with a weight factor $1/\cos^{1-2\ep}\theta$
(see eqs.~(3.38)--(3.39) of \cite{DD}).
This triangle 123 is split into three triangles 012, 023
and 031. Then, each of them is split into two
rectangular triangles, according to Fig.~9 of \cite{DD}.
We consider
the contribution of one of the six resulting triangles,
namely the left rectangular triangle in Fig.~9.
Its angle at the vertex~0 is denoted as 
${\textstyle{1\over2}}\varphi_{12}^{+}$, whereas the
height dropped from the vertex~0 is denoted $\eta_{12}$.
 
As compared to the four-dimensional
($\ep=0$) case described in section~VB of \cite{DD},
eq.~(5.14) should be transformed (for non-zero $\ep$) into
\be
\label{5.14+}
-\frac{1}{2\ep}\;\frac{\partial}{\partial\xi}
\cos^{2\ep}\theta(\xi,\varphi) ,  
\ee
where $\xi$ is an auxiliary variable ($0\leq\xi\leq1$).
Integrating (\ref{5.14+}) over $\xi$ yields
\be
\frac{1}{2\ep} \left[ \cos^{2\ep}\theta(0,\varphi)
- \cos^{2\ep}\theta(1,\varphi) \right]
= \frac{1}{2\ep} \left[ 1 - 
\left( 1 + \frac{\tan^2\eta_{12}}{\cos^2\varphi} \right)^{-\ep} \right] .
\ee
Therefore, the remaining $\varphi$ integration is (cf.\ eq.~(5.16) of
\cite{DD})
\bea
\label{last}
\frac{1}{2\ep} \int_0^{\varphi_{12}^{+}/2} \mbox{d}\varphi 
\left[ 1 -
\left( 1 + \frac{\tan^2\eta_{12}}{\cos^2\varphi} \right)^{-\ep} \right] 
\hspace{45mm}
\nn \\
= \frac{1}{2} \sum\limits_{j=0}^{\infty}\frac{(-\ep)^j}{(j+1)!}\;
\int_0^{\varphi_{12}^{+}/2} \mbox{d}\varphi
\ln^{j+1}\left( 1 + \frac{\tan^2\eta_{12}}{\cos^2\varphi} \right) .
\eea
In the limit $\ep\to0$ 
we get a combination
of $\mbox{Cl}_2$ functions, eq.~(5.17) of \cite{DD}.

Collecting the results for all six triangles,
we get the result for the three-point function with 
arbitrary masses and external momenta, corresponding
(at $\ep=0$) to the analytic continuation of the 
well-known formula presented in \cite{`tHV-79}. 
The $\ep$-term of the three-point function has been 
calculated in \cite{NMB} in terms of 
$\mbox{Li}_3$. The case of massless internal particles 
has been considered in \cite{UD3}. Its analytic continuation
in terms of $\mbox{Ls}_3$ functions has been constructed 
in \cite{DT2}. Note that the same functions occur in the
results for the two-loop vacuum diagrams with different
masses. This connection was observed in \cite{DT1} and
then explained in \cite{DT2}.
We believe that the higher terms of the expansion 
(\ref{last}) can also be presented in a compact form
in terms of the $\mbox{Ls}_j$ functions.

\vspace{3mm}

$\hspace*{-2mm}$
{\bf Acknowledgements.} I am grateful to R.~Delbourgo, 
M.Yu.~Kalmykov, O.V.~Tarasov and J.B.~Tausk for useful discussions.
My research and participation in the AIHENP-99 conference
were supported by the Alexander von Humboldt Foundation.
I am grateful to the organizers for their hospitality.

\newpage


\begin{thebibliography}{99}

\bibitem{Landau}
L.D.~Landau, {\em Nucl.~Phys.} {\bf 13} (1959) 181.

\bibitem{KW}
G.~K\"allen and A.~Wightman,
{\em Mat.~Fys.~Skr.~Dan.~Vid.~Selsk.} {\bf 1}, No.6 (1958) 1.

\bibitem{Wu}
A.C.T.~Wu, {\em Mat.~Fys.~Medd.~Dan.~Vid.~Selsk.}
             {\bf 33}, No.3 (1961) 1.

\bibitem{DD}
A.I.~Davydychev and R.~Delbourgo, 
{\em J.~Math.~Phys.} {\bf 39} (1998) 4299.

\bibitem{DD-APP}
A.I.~Davydychev and R.~Delbourgo, 
{\em Acta~Phys.~Pol.} {\bf B29} (1998) 2891.

\bibitem{Nickel}
B.G.~Nickel, {\em J.~Math.~Phys.} {\bf 19} (1978) 542.

\bibitem{Wagner}
N.~Ortner and P.~Wagner,
{\em Ann.~Inst.~Henri~Poincar\'e (Phys.~th\'eor.)}
 {\bf 63} (1995) 81;\\
P.~Wagner, {\em Indag.~Math.} {\bf 7} (1996) 527.

\bibitem{BB}
J.M.~Borwein and D.J.~Broadhurst, preprint OUT-4102-76, Milton Keynes,
    1998 (hep-th/9811173).

\bibitem{dimreg}
G.~'tHooft and M.~Veltman,
{\em Nucl.~Phys.} {\bf B44} (1972) 189;\\
C.G.~Bollini and J.J.~Giambiagi,
{\em Nuovo~Cimento} {\bf 12B} (1972) 20; \\
J.~F.~Ashmore,
{\em Lett.~Nuovo~Cim.} {\bf 4} (1972) 289;\\ 
G.~M.~Cicuta and E.~Montaldi,
{\em Lett.~Nuovo~Cim.} {\bf 4} (1972) 329.

\bibitem{BDS}
F.A.~Berends, A.I.~Davydychev and V.A.~Smirnov,
{\em Nucl.~Phys.} {\bf B478} (1996) 59.

\bibitem{Lewin}
L.~Lewin, {\em Polylogarithms and associated functions},
      North Holland, 1981.

\bibitem{DT2}
A.I.~Davydychev and J.B.~Tausk, 
{\em Phys.~Rev.} {\bf D53} (1996) 7381.

\bibitem{NMB}
U.~Nierste, D.~M\"uller and M. B\"ohm, {\em Z.~Phys.}
      {\bf C57} (1993) 605.

\bibitem{FJTV}
J.~Fleischer, F.~Jegerlehner, O.V.~Tarasov and O.L. Veretin,
{\em Nucl.~Phys.} {\bf B539} (1999) 671.

\bibitem{FKK}
J.~Fleischer, M.Yu.~Kalmykov and A.V.~Kotikov, 
hep-ph/9905249; \\
J.~Fleischer and M.Yu.~Kalmykov, hep-ph/9907431.

\bibitem{`tHV-79} G.~'tHooft and M.~Veltman, {\em Nucl.~Phys.}
             {\bf B153} (1979) 365.

\bibitem{UD3}
N.I.~Ussyukina and A.I.~Davydychev, 
{\em Phys.~Lett.} {\bf B332} (1994) 159.

\bibitem{DT1}
A.I.~Davydychev and J.B.~Tausk, 
{\em Nucl.~Phys.} {\bf B397} (1993) 123.

\end{thebibliography}
\end{document}